\documentstyle[aps,prbbib,twocolumn,epsf]{revtex}
\begin{document}
\draft
\title{Quantization of adiabatic pumped charge
in the presence of superconducting lead}

\author{Jian Wang and Baigeng Wang}

\address{Department of Physics, The University of Hong Kong, 
Pokfulam Road, Hong Kong, China}
\maketitle

\begin{abstract}
We investigate the parametric electron pumping of a double barrier
structure in the presence of a superconducting lead. The parametric
pumping is facilitated by cyclic variation of the barrier heights $x_1$
and $x_2$ of the barriers. In the weak coupling regime, there exists
a resonance line in the parameter space $(x_1,x_2)$ so that the energy
of the 
quasi-bound state is in line with the incoming Fermi energy. Levinson
et al found recently that the pumped charge for each pumping cycle is 
quantized with $Q=2e$ for normal structure when the pumping contour 
encircles the resonance line. In the presence of a superconducting lead, 
we find that the pumped charge is quantized with the value $2e$.
\end{abstract}


Physics of parametric electron pump has attracted great attention
recently\cite{brouwer,switkes,zhou,wagner,Avron,aleiner1,wei1,vavilov,brouwer2,sharma,kravtsov,wei2}.
A classical example of electron pump is the Thouless pump facilitated 
by a traveling wave potential\cite{thouless}. The pumped charge is 
quantized\cite{thouless} and can be used as a quantum standard for  
electric charge.\cite{niu}.
The quantization of pumped charge has also been studied for a large, 
almost open quantum dot\cite{aleiner,shutenko} and a small, strongly 
pinched quantum dot\cite{levinson}. In the latter case, there exists
a resonance line along which the transmission through the quantum dot
is at resonance. The pumped charge is quantized if the pumping contour 
in parameter space is properly chosen to encircle the resonance
line\cite{levinson}. Recently, we have studied the parametric pumping in 
presence of a superconducting lead\cite{apl}. At the normal 
conductor-superconductor 
(NS) interface, an incoming electron-like excitation can be Andreev 
reflected as a hole-like excitation\cite{andreev}. In contrast to the 
current doubling effect\cite{beenakker}, we found that due to the quantum 
interference of direct reflection and the multiple Andreev reflection, 
the pumped current is four times of the value when the leads are normal 
in the {\it weak pumping regime}. In this paper, we explore the effect of 
superconducting lead on electron pumping in the opposite limit, i.e., we 
study the pumped charge during the pumping cycle in the the strong pumping 
regime. Here the pumped charge equal to the pumped current multiplied by 
the period of pumping cycle. Similar to the Ref.\onlinecite{levinson}, we 
examine the behavior of pumped charge near the resonance line. We find 
that the pumped charge in one pumping cycle is quantized with the 
value of $Q=2e$ when one of the leads is superconducting.\cite{foot2} 

We consider a parametric pump which consists of a double barrier 
tunneling structure attached to a normal left lead and a
superconducting right lead. 
Due to the cyclic variation of external parameters $x_1$ and $x_2$,
the adiabatic charge transfer in the presence of 
a superconducting lead is\cite{brouwer,foot4,foot1}
\begin{equation}
Q^{NS}=2e\int_0^\tau dt [\frac{dN_L}{dx_1} 
\frac{dx_1}{dt} + \frac{dN_L}{dx_2} \frac{dx_2}{dt}] 
\label{q1}
\end{equation} 
where $\tau$ is the period of cyclic variation and the quantity $dN_L/dx$ 
is the injectivity\cite{buttiker,jwang} given, at zero temperature, by
\begin{eqnarray}
\frac{dN_L}{dx_j}= \frac{1}{2\pi} Im [{\cal S}^*_{ee} 
\frac{\partial {\cal S}_{ee}}{\partial x_j} 
- {\cal S}^*_{he} \frac{\partial {\cal S}_{he}}{ \partial x_j}] 
\label{inj}
\end{eqnarray}
where the first term is the injectivity of electron due to the
variation of the external parameter\cite{buttiker,jwang}, {\it i.e.} the 
partial density of states (DOS) for an electron coming from left lead 
and exiting the system as an electron,
and the second term is the injectivity of a hole, 
{\it i.e.} the DOS for a hole coming from left lead and exiting the
system as an electron. Using the Green's theorem, the pumped charge
can be expressed as surface integral over area A enclosed
by the path $(x_1(t), x_2(t))$ in the parameter space\cite{brouwer}
\begin{equation}
Q^{NS}=\frac{2e}{\pi}\int_A dx_1 dx_2 \Pi^{NS}(x_1,x_2)
\label{q2}
\end{equation} 
with 
\begin{equation}
\Pi^{NS}(x_1,x_2)=
Im [\frac{\partial {\cal S}^*_{ee}} 
{\partial x_1} \frac{\partial {\cal S}_{ee}}{\partial x_2} 
- \frac{\partial {\cal S}^*_{he}}{\partial x_1} \frac{\partial 
{\cal S}_{he}}{\partial x_2}] 
\end{equation} 
Note that the area $A$ is a measure of variation of pumping parameters
$x_1$ and $x_2$. $A$ is very small in the weak pumping limit while
remains finite in the strong pumping regime.

For the NS structures, the scattering matrix is described by
$2\times 2$ matrix $\hat{{\cal S}}$ when the Fermi energy is within
the superconducting gap $\Delta$.

\begin{equation}
\hat{{\cal S}} = \left( \begin{array}{ll}
         {\cal S}_{ee} & {\cal S}_{eh} \\
         {\cal S}_{he} & {\cal S}_{hh}
         \end{array}
  \right)
\end{equation}
where ${\cal S}_{ee}$ (or ${\cal S}_{he}$) is the scattering
amplitude of the incident electron reflected as an electron (or a
hole). Using Andreev approximation\cite{andreev}, we 
have\cite{beenakker,lesovik}
\begin{equation}
\hat{{\cal S}} = \hat{S}_{11} + \hat{S}_{12} (1 - \hat{R}_I
\hat{S}_{22})^{-1} \hat{R}_I \hat{S}_{21}
\label{lesovik}
\end{equation}
where $\hat{S}_{\beta \gamma}(E)$ ($\beta, \gamma=1,2$) is a diagonal 
$2 \times 2$ scattering matrix for the double barrier structure with
matrix element $S_{\beta \gamma}(E)$ and $S^*_{\beta \gamma}(-E)$. 
For instance, we have

\begin{equation}
\hat{S}_{11} = \left( \begin{array}{ll}
         S_{11}(E) & ~~~ 0 \\
         0 & ~~~ S^*_{11}(-E)
         \end{array}
  \right)
\end{equation}
In Eq.(\ref{lesovik}) $\hat{R}_I=\alpha \sigma_x$ is the $2\times 2$ 
scattering matrix at NS interface due to the Andreev reflection
with off diagonal matrix element 
$\alpha$. Here $\alpha = (E-i\nu \sqrt{\Delta^2-E^2})/\Delta$ with 
$\nu=1$ when $E>-\Delta$ and $\nu=-1$ when $E<-\Delta$. In
Eq.(\ref{lesovik}), the energy $E$ is measured relative to the chemical
potential $\mu$ of the superconducting lead.  Eq.(\ref{lesovik}) 
has a clear physical meaning\cite{lesovik}. The first term is the
direct reflection from the normal scattering structure and the second
term can be expanded as $\hat{S}_{12} \hat{R}_I \hat{S}_{21} + \hat{S}_{12} 
\hat{R}_I \hat{S}_{22} \hat{R}_I \hat{S}_{21} + ...$ which is
clearly the sum of the multiple Andreev reflection in the hybrid structure.
It is the quantum interference of these two terms which gives rise
the enhancement of pumped current in the weak pumping regime for NS
system\cite{apl}. From Eq.(\ref{lesovik}) we obtain the well known 
expressions for 
the scattering matrix ${\cal S}_{ee}$ and ${\cal S}_{he}$\cite{beenakker}
\begin{equation}
{\cal S}_{ee}(E) = S_{11}(E) + \alpha^2 S_{12}(E) S_{22}^*(-E) M_e 
S_{21}(E)
\label{see}
\end{equation}

\begin{equation}
{\cal S}_{he}(E) = \alpha S_{12}^*(-E) M_e S_{21}(E)
\label{she}
\end{equation}
and 
\begin{equation}
M_e = [1- \alpha^2 S_{22}(E) S_{22}^*(-E)]^{-1}
\label{me}
\end{equation}
The double barrier structure which we consider is modeled by potential 
$U(y)=V_1 \delta (y+a/2)+V_2 \delta (y-a/2)$ where $V_1$ and $V_2$ are 
barrier heights which varys in a cyclic fashion to allow the charge
pumping. For this system the retarded Green's function $G^r(y,y')$ can be 
calculated exactly\cite{yip}. This is done by applying the Dyson's equation 
regarding that any one of the $\delta$-barrier is just a perturbation of 
the remaining system. This way $G^r(y,y')$ is obtained by applying 
Dyson's equation twice starting from the Green's function of the 
one-dimensional free space.  With $G^r(y,y')$ we can calculate 
scattering matrix exactly from the Fisher-Lee relation\cite{lee}
\begin{equation}
S_{\beta \gamma}=-\delta_{\beta \gamma}+i v G^r_{\beta \gamma} 
\label{fisher}
\end{equation}
where $G^r_{\beta \gamma} = G^r(y_\beta,y_\gamma)$ and $v=2k$ is the 
electron velocity in the normal lead. For normal structure, we 
have\cite{levinson}
\begin{equation}
S_{11} = [1-i x_2 - (1+i x_1) \sigma^2]/D
\end{equation}
\begin{equation}
S_{22} = [1-i x_1 - (1+i x_2) \sigma^2]/D
\end{equation}
and 
\begin{equation}
S_{12} = S_{21} = x_1 x_2 \sigma /D
\end{equation}
where $D = -(1-i x_1)(1-i x_2)+\sigma^2$, $x_{1,2}=2k V_{1,2}$, and 
$\sigma=\exp(ika)$. For the double barrier structure, the resonant
tunneling is mediated by the quasi-bound state. When the energy of the
incident electron is in line with the energy of the quasi-bound
state the transmission coefficient reaches maximum. The energy of
quasi-bound states can be determined either by looking at the pole of
the scattering matrix\cite{levinson} which works well in one
dimension or by calculating the dwell time of the incident 
electron for two or three dimensional systems\cite{apl1}. In the case
of double $\delta$ barriers structure, the energy of
quasi-bound state is given by\cite{levinson} $E = E_r + \Delta E$
with $\Delta E = -(k_r/a)(x_1+x_2)$ where $E_r = k_r^2 = (n\pi/a)^2$ 
is energy of the bound state when the system is isolated. This
defines a resonance line $x_1+x_2=-\delta$ in parameter space 
$(x_1,x_2)$ along which the transmission is at 
resonance\cite{levinson}. Here $\delta<0$ is the detuning of the Fermi 
energy from the bound state.

To show the quantization of charge transfer in the NS system, it is
useful to recall the calculation of the normal case and make the
comparison. In the normal case the charge transfer is given
by\cite{brouwer,foot2}
\begin{equation}
Q^{N}=\frac{2e}{\pi}\int_A dx_1 dx_2 \Pi^{N}(x_1,x_2)
\label{normal}
\end{equation} 

\begin{equation}
\Pi^{N}(x_1,x_2)=
Im [\frac{\partial {\cal S}^*_{11}} 
{\partial x_1} \frac{\partial {\cal S}_{11}}{\partial x_2} 
+ \frac{\partial {\cal S}^*_{12}}{\partial x_1} \frac{\partial 
{\cal S}_{12}}{\partial x_2}] 
\label{normal1}
\end{equation} 
The pumped charge in this case has been calculated in
Ref.\onlinecite{levinson}. In the {\it weak pumping limit}, it is 
easy to show that only $\partial_x S_{11}$ contributes to the 
pumped charge. In the strong pumping regime, we will show in the 
following that the contribution from $\partial_x S_{12}$ to 
the pumped charge in normal structure is zero. 
As discussed in detail in Ref.\onlinecite{levinson}, we neglect
the smooth energy dependence of $x_1$ and $x_2$.
From Eq.(\ref{normal1}), we obtain the contribution due to 
$\partial_x S_{12}$
\begin{equation}
\Pi_1^{N}(x_1,x_2)=F_1(x_1,x_2)/F_2^2(x_1,x_2)
\label{pi1}
\end{equation}
with
\begin{equation}
F_1(x_1,x_2)= -2x_1 x_2 (x_1-x_2) \sin^2(\delta/2)
\label{f1}
\end{equation}
\begin{eqnarray}
F_2(x_1,x_2)&=&x_1^2 x_2^2 +(x_1+x_2)^2 +2(x_1+x_2)\sin\delta
\nonumber \\
&+&2(1-x_1x_2)(1-\cos\delta)
\label{f2}
\end{eqnarray}
To compute the surface integral of $\Pi^N_1$ in Eq.(\ref{pi1}), it
is convenient to change the variables from $x_{1,2}$ to $p$ and 
$z$: 

\begin{equation}
x_1 = -p ~ \delta ~ (1+z)/2 
\label{x1}
\end{equation}
and
\begin{equation} 
x_2 = -p~ \delta ~ (1-z)/2 
\label{x2}
\end{equation}
with $0<p<\infty$ and $-1<z<1$. Substituting Eqs.(\ref{x1}) and 
(\ref{x2}) into Eqs.(\ref{f1}) and (\ref{f2}) 
and expanding Eqs.(\ref{f1}) and (\ref{f2}) in terms of small 
$\delta$, we have
\begin{equation}
F_1=z(1-z^2)\delta^5 p^3/8
\end{equation}
\begin{equation}
F_2=\delta^2[(1-p)^2+\delta^2 g(p,z)]
\end{equation}
where $g(p,z)$ (an even function of $z$) is given in Eq.(8) of 
Ref.\onlinecite{levinson}. Since $F_1$ is an odd function of $z$, the
contribution due to $\partial_x S_{12}$ to the pumped charge is zero.

Now we follow the same procedure to calculate the pumped charge for the
NS system. For the parametric pumping at zero temperature, we only need 
the scattering matrix at the Fermi level, i.e., at $E=0$. From
Eqs.(\ref{she}) and (\ref{me}), we see that ${\cal S}_{he}$ is a real
quantity and hence makes no contribution to the pumped charge in
Eq.(\ref{q2}). It is straightforward to calculate $\Pi^{NS}$ using
Eq.(\ref{see}), from which we obtain,
$\Pi^{NS}(x_1,x_2)=F_3(x_1,x_2)/F_4^3(x_1,x_2)$
where
$F_3 = 4x_1^4 x_2^3 (2-2\cos\delta+x_2\sin\delta)$
and 
$F_4 = x_1^2 x_2^2 +2(x_1+x_2)^2+4(x_1+x_2)\sin\delta 
+4(1-x_1x_2)(1-\cos\delta)$. 

In Fig.1 we plot both $\Pi^{NS}$ and $\Pi^N$ as well as their 
cross-sections 
along and perpendicular to the resonance line. We see that $\Pi^{NS}$
and $\Pi^N$ are peaked around the resonance line. Two features are worth 
noticing. First of all, the peak of $\Pi^{NS}$ is much sharper than that 
of $\Pi^N$. This is understandable and is due to the resonance nature 
of NS structures near the resonance line. In the Breit-Wigner form, the 
transmission coefficients for normal and NS structures are, respectively
$|S_{21}|^2 = \Gamma_1 \Gamma_2/[(E-E_r)^2+\Gamma^2/4]$
and\cite{beenakker} 
$|{\cal S}_{he}|^2 = 4\Gamma_1^2 \Gamma_2^2/[4(E-E_r)^2
+\Gamma_1^2+\Gamma_2^2]^2$, 
where $E_r$ is the resonant level, $\Gamma_1$ and $\Gamma_2$ are
the decay widths into the left and right lead. 
Hence $|{\cal S}_{he}|^2$ decays much faster away from $E_r$ than 
$|S_{21}|^2$. The scattering matrix $S_{21}$ and ${\cal S}_{he}$ will 
appear, respectively, in Eqs.(\ref{q2}) and (\ref{normal}) implicitly as 
can be seen from Fisher-Lee relation Eq.(\ref{fisher}) and the Dyson 
equation $\partial_{X_2} G^r_{11} = G^r_{12} G^r_{21}$\cite{gasparian}. 
Secondly, the peak height of $\Pi^{NS}$ is four times larger than that of 
$\Pi^N$.  This is precisely due to the constructive interference of direct 
reflection and multiple Andreev reflection\cite{apl}. 
Now the physics of pumping at resonance is clear. For the resonance 
pumping in the weak pumping regime, we are looking at the small 
neighborhood of the peak. The area of the neighborhood 
has to be small since it is the weak pumping. The neighborhood has to be 
around the peak with $x_1 \sim x_2$ since only around the peak the 
transmission coefficient is approximately one. As a result, we obtain 
immediately the pumped charge or pumped current of NS structure near 
the resonance is four times of that of corresponding normal structure. 
In the other extreme, for strong pumping, we take a large contour 
enclosing entire resonance line. Since $\Pi^{NS}$ decreases much 
faster than $\Pi^N$ away from the peak, it is understandable that the 
pumped charges (the integral of $\Pi$ over the area enclosed by the 
contour) for both normal and NS structures are equal, which will be 
shown analytically below. 

After the expansion in powers of $\delta$ in Eqs.(\ref{f1}) and 
(\ref{f2}) and keep the leading orders of $\delta$, we have
\begin{equation}
F_3=p^7[2+p(-1+z)](-1+z)^3 (1+z)^4\frac{\delta^9}{64}
\end{equation}

\begin{eqnarray}
F_4&=&2(1-p)^2 \delta^2 + [-\frac{1}{6}+\frac{2p}{3}
\nonumber \\
&+&\frac{1}{2}p^2(-1+z^2) +\frac{1}{16}p^4(-1+z^2)^2] \delta^4
\end{eqnarray}

So Eq.(\ref{q2}) becomes,
\begin{equation}
Q^{NS} = \frac{e}{\pi} \int_0^{\infty} p dp \int_{-1}^1 dz 
\frac{F_3} {F_4^3} ~ \delta^2
\label{eq1}
\end{equation}
using the fact that $\lim_{\delta->0} \delta^5/(x^2+\delta^2)^3 = 
(3/8)\pi \delta(x)$, Eq.(\ref{eq1}) becomes

\begin{equation}
Q^{NS} = 3\sqrt{2}e\int_{-1}^1 dz \frac{(1-z^2)^3(1+z)^2}{(1
+6z^2+z^4)^{5/2}} = 2e
\end{equation}
Hence the pumped charge for NS system is quantized at the same 
value of that of the normal structure.

Now we have a better physical picture for the transport properties
of the NS structure. For the conductance or the I-V curve, we need 
$S_{21}$ or ${\cal S}_{he}$. For normal structure, the current is 
given by $I^N = 2e/h \int dE [f(E-eV_1) - f(E-eV_2)] |S_{12}|^2$
and hence at resonance and at zero temperature $G^N = I^N/(V_1-V_2) 
= 2e^2/h$. For NS structure, we
have\cite{beenakker,sun} $I^{NS} = 2e/h \int dE 
(f(E-eV_1)-f(E+eV_1)) |{\cal S}_{he}|^2$ and at resonance $G^{NS} =
I^{NS}/V_1 = 4e^2/h$ which is the well known doubling of the conductance. 
For pumped charge or pumped current at resonance, however, it depends
only on $\partial_{x_i} S_{11}$ or $\partial_{x_i} {\cal S}_{ee}$ ($E=0$
is assumed). Because of the constructive interference between direct 
reflection and multiple Andreev reflection in the {\it weak pumping 
regime}, the charge transfer increases by a factor of four when one of 
the lead becomes superconducting. In the strong pumping regime, 
however, the charge transfer 
is quantized at the value equal to that of normal structure, if the 
pumping contour is chosen such that the resonance line is enclosed. 
The physics behind this can be understood as follows. In the normal
case, the contour enclosing the resonance line in the parameter space 
passes through the resonance line at two points $(x_1,x_2)=
(0,-\delta)$ when the left contact is almost closed and $(-\delta,0)$
when the right contact is almost closed. When passing through those
two points, the resonance level of the dot crosses the Fermi
energy.  At each crossing, the occupation of the level changes, and 
two electrons with opposite spin enter or exit the region between the 
barriers. Since one of the tunnel barriers has zero
conductance at those points, it is clear that the electrons must
have tunneled through the other contact upon entering or leaving
the quantum dot. Hence, in the pumping cycle, electrons are
shuttled pairwise through the dot. In the presence of
superconducting lead, the resonance level (both the energy and the
width) is exactly the same as that of normal case since the scattering 
matrix is given by ${\cal S}_{he} = i |S_{12}|^2/(1+|S_{22}|^2)$ 
when $E=0$. Therefore the same argument applies to the
superconducting case and the quantization unit is $2e$. Note that our
statement is only valid when the electron interaction is neglected.
For the case of two normal-metal contacts, if interactions are included 
the quantization will remain, but now the quantum is only $e$: Only one 
electron at a time can enter the region between the barriers; addition of a
second electron is forbidden by Coulomb blockade. In the presence of
superconducting lead, since the Andreev reflection requires two
electrons with opposite spin in order to produce the supercurrent, it 
seems that the pumping is not allowed in the strong pumping regime due 
to Coulomb blockade. 
In this paper, we have also neglected the effect of temperature and the 
effect of inelastic scattering. As discussed in Ref.\onlinecite{levinson}
the temperature will destroy the quantization of the pumped charge. When
inelastic channel is present an additional physical mechanism for an 
incoherent pump effect will show up\cite{buttiker1}.

\section*{Acknowledgments}
We gratefully acknowledge support by a RGC grant from the SAR Government of 
Hong Kong under grant number HKU 7215/99P.

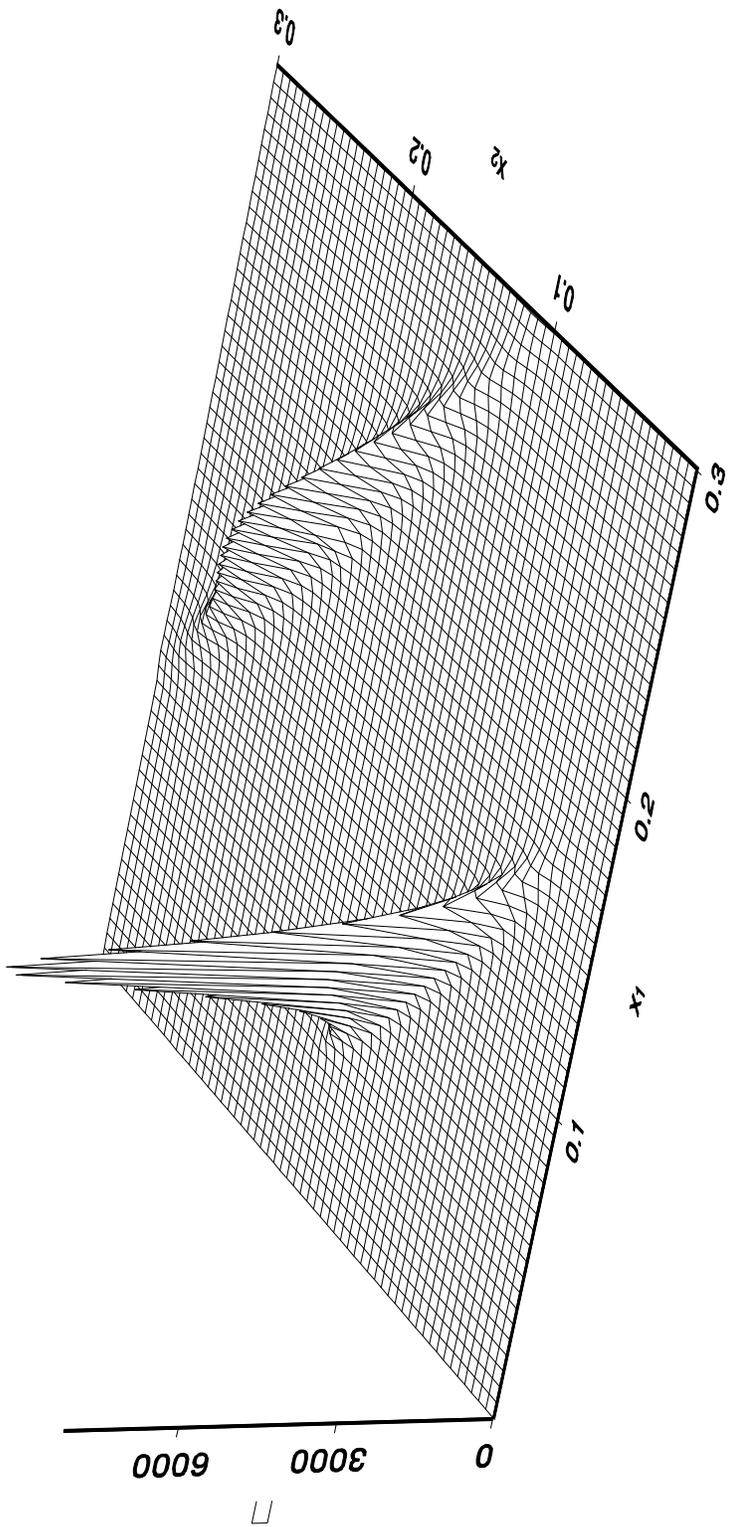
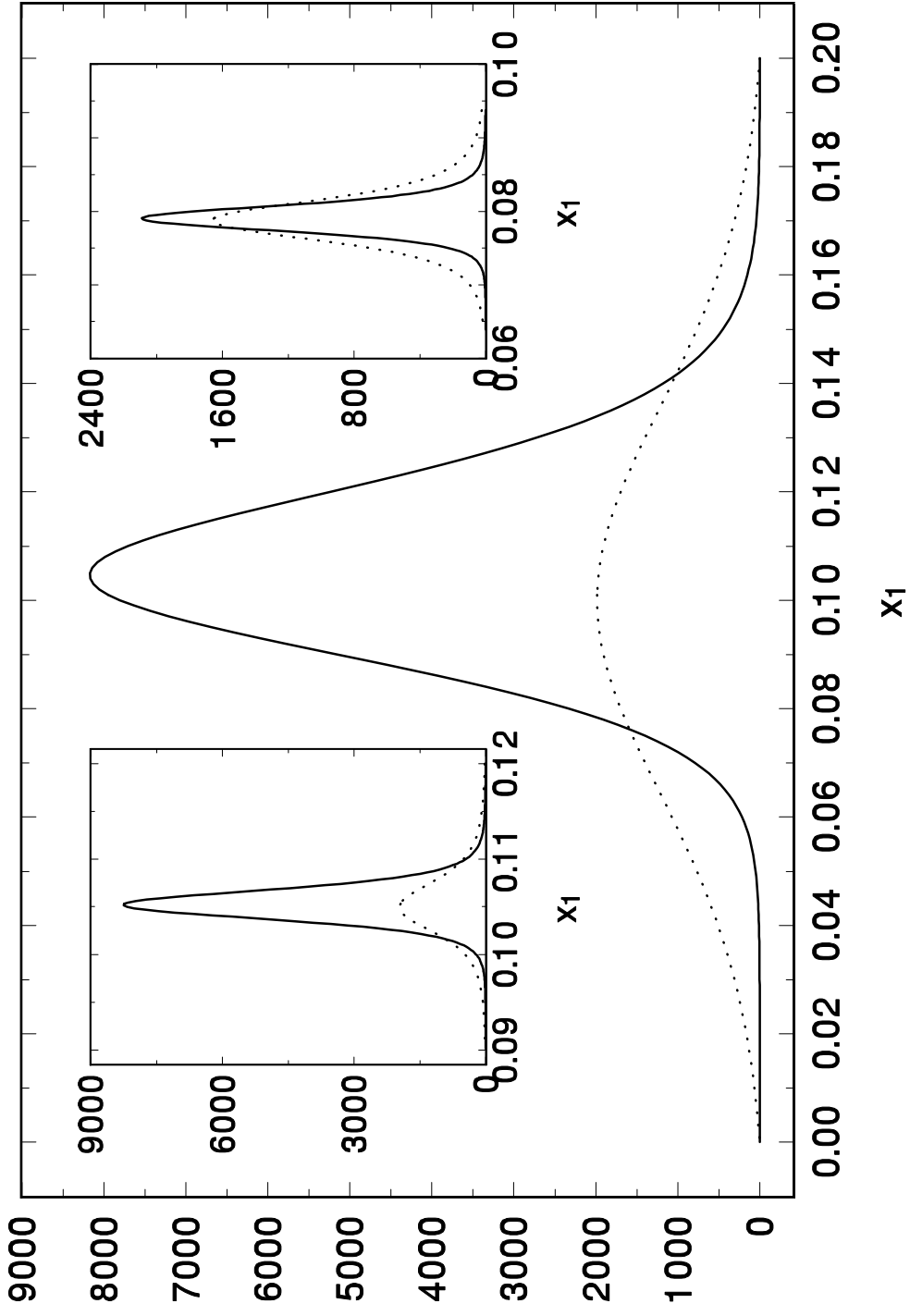
\begin{figure}
\caption{
(a). The integrand $\Pi$ of Eqs.(3) and (15) as a function of $x_1$
and $x_2$ for $\delta=-0.2$.  For illustrating purpose, the origin 
of $\Pi^N(x_1,x_2)$ has been shifted by $(0.1,0.1)$. (b). The 
cross-section of $\Pi$ along the resonance line $x_1+x_2=-\delta$. 
Solid line: $\Pi^{NS}$; dotted line: $\Pi^N$. Inset: the cross-section
of $\Pi$ along the direction $x_1-x_2=c_0$ which is perpendicular to 
the resonance line. Left inset: $c_0=0.01$; right inset: $c_0=-0.042$.
}
\end{figure}


\medskip

\begin{thebibliography}{00} 

\bibitem{brouwer}
P.W. Brouwer, Phys. Rev. B {\bf 58}, R10135 (1998). 

\bibitem{switkes}
M. Switkes, C. Marcus, K. Capman, and A.C. Gossard, Science {\bf 283}, 
1905 (1999). 

\bibitem{zhou}
F. Zhou, B. Spivak, and B.L. Altshuler, Phys. Rev. Lett. {\bf 82}, 
608 (1999).

\bibitem{wagner}
M. Wagner, Phys. Rev. Lett. {\bf 85}, 174 (2000).

\bibitem{Avron}
J.E. Avron, A. Elgart, G.M. Graf, and L. Sadun, Phys. Rev. B {\bf
62}, R10618 (2000).

\bibitem{aleiner1}
I.L. Aleiner, B.L. Altshuler, and A. Kamenev, Phys. Rev. B {\bf 62},
10373 (2000). 

\bibitem{wei1}
Y.D. Wei, J. Wang, and H. Guo, Phys. Rev. B {\bf 62}, 9947 (2000). 

\bibitem{vavilov}
M.G. Vavilov, V. Ambegaokar, and I.L. Aleiner, Phys. Rev. B {\bf
63}, 195313 (2001). 

\bibitem{brouwer2}
P.W. Brouwer, Phys. Rev. B {\bf 63}, 121303 (2001); 
M.L. Polianski and P.W. Brouwer, Phys. Rev. B {\bf 64}, 075304
(2001).

\bibitem{sharma}
P. Sharma and C. Chamon, Phys. Rev. Lett. {\bf 87}, 096401 (2001)

\bibitem{kravtsov}
X.B. Wang and V.E. Kravtsov, Phys. Rev. B {\bf 64}, 033313 (2001).

\bibitem{wei2}
Y.D. Wei, J. Wang, H. Guo, and C. Roland, Phys. Rev. B {\bf 64}, 
115321 (2001). 

\bibitem{thouless}
D.J. Thouless, Phys. Rev. B {\bf 27}, 6083 (1983). 

\bibitem{niu}
Q. Niu, Phys. Rev. Lett. {\bf 64}, 1812 (1990).

\bibitem{aleiner}
I.L. Aleiner and A.V. Andreev, Phys. Rev. Lett. {\bf 81}, 1286 (1998). 

\bibitem{shutenko}
T.A. Shutenko, I.L. Aleiner, and B.L. Altshuler, Phys. Rev. B {\bf
61}, 10366 (2000). 

\bibitem{levinson}
Y. Levinson, O. Entin-Wohlman, and P. Wolfle, cond-mat/0010494.

\bibitem{apl}
J. Wang et al, Appl. Phys. Lett. {\bf
79}, 3977 (2001). 

\bibitem{andreev}
A.F. Andreev, Zh. Eksp. Teor. Fiz. {\bf 46}, 1823 (1964) [Sov. Phys.
JETP {\bf 19}, 1228 (1964)].

\bibitem{beenakker}
C.W.J. Beenakker, Rev. Mod. Phys. {\bf 69}, 731 (1997).

\bibitem{foot2}
A factor of two has been included for spin.


\bibitem{foot4}
The units are fixed by setting $\hbar=2m=1$ in the following 
analysis.  For the GaAs system with $a=1000A$, the energy uint is 
$E=56 \mu eV$. 

\bibitem{foot1}
This formula can be derived using the Keldysh nonequilibrium Green's 
function method (see Ref.\onlinecite{wbg1}) and we have assumed that 
the Coulomb blockade effect can be neglected.

\bibitem{wbg1}
B.G. Wang, J. Wang, and H. Guo, cond-mat/0107078.

\bibitem{buttiker}
T. Gramespacher and M. Buttiker, Phys. Rev. B {\bf 61}, 8125 (2000).

\bibitem{jwang}
J. Wang, Y.D. Wei, H. Guo, Q.F. Sun, and T.H. Lin, Phys. Rev. B {\bf
64}, 104508 (2001). 

\bibitem{lesovik}
G.B. Lesovik, A.L. Fauchere, and G. Blatter, Phys. Rev. B {\bf 55},
3146 (1997). 

\bibitem{lee}
D. S. Fisher and P. A. Lee, Phys. Rev. B {\bf 23}, 6851 (1981). 

\bibitem{apl1}
J. Wang, Y.J. Wang, and H. Guo, Appl. Phys. Lett. {\bf 65}, 1793
(1994). 

\bibitem{yip}
M.K. Yip, J. Wang, and H. Guo, 
Z. Phys. B: Condens. Matter {\bf 104}, 463 (1997). 

\bibitem{gasparian}
V. Gasparian, T. Christen, and M. Buttiker, Phys. Rev. A {\bf 54}, 4022 
(1996). 

\bibitem{sun}
Q.F. Sun, J. Wang, and T.H. Lin, Phys. Rev. B {\bf 59}, 3831 (1999).

\bibitem{buttiker1}
M. Moskalets and M. Buttiker, cond-mat/0108061.

\end{thebibliography}
\end{document}